\documentclass{article}
\usepackage{authblk}

\usepackage[utf8]{inputenc} 
\usepackage[T1]{fontenc}    
\usepackage{hyperref}       
\usepackage{url}            
\usepackage{booktabs}       
\usepackage{amsfonts}       
\usepackage{nicefrac}       
\usepackage{microtype}      
\usepackage{fullpage}
\usepackage{enumerate}
\usepackage{enumitem}
\usepackage{amsmath}
\usepackage{amssymb}
\usepackage{algorithmicx}
\usepackage{algorithm}
\usepackage{algpseudocode}
\usepackage{mathtools}
\usepackage{graphicx}

\usepackage{sidecap}
\newcommand{\RR}{\mathbb{R}}
\newcommand{\CC}{\mathbb{C}}
\newcommand{\ZZ}{\mathbb{Z}}

\newcommand{\M}{\mathbf{M}}

\newcommand{\G}{\mathbf{G}}

\newcommand{\T}{\mathbf{T}}

\newcommand{\B}{\mathbf{B}}

\newcommand{\RS}[2]{\ifthenelse{\equal{#1}{}}{\mathsf{RS}}{\mathsf{RS}[#1,#2]}}

\newcommand{\floor}[1]{\left\lfloor#1\right\rfloor}
\newcommand{\ceil}[1]{\left\lceil#1\right\rceil}

\newcommand{\card}[1]{{|#1|}}

\renewcommand{\t}{\mathsf{T}}

\newtheorem{prop}{Proposition}

\newtheorem{defn}{Definition}
\newtheorem{clry}{Corollary}
\newtheorem{lem}{Lemma}

\newtheorem{rem}{Remark}

\title{Improving Distributed Gradient Descent Using Reed--Solomon Codes}

\author{Wael Halbawi}
\author{Navid Azizan-Ruhi}
\author{Fariborz Salehi}
\author{Babak Hassibi}
\affil{Department of Electrical Engineering\\
California Institute of Technology\\
Pasadena, CA 91125}

\begin{document}

\maketitle

\begin{abstract}
Today's massively-sized datasets have made it necessary to often perform computations on them in a distributed manner. In principle, a computational task is divided into subtasks which are distributed over a cluster operated by a taskmaster. One issue faced in practice is the delay incurred due to the presence of slow machines, known as \emph{stragglers}. Several schemes, including those based on replication, have been proposed in the literature to mitigate the effects of stragglers and more recently, those inspired by coding theory have begun to gain traction. In this work, we consider a distributed gradient descent setting suitable for a wide class of machine learning problems. We adapt the framework of Tandon et al. \cite{Tandon2016} and present a deterministic scheme that, for a prescribed per-machine computational effort, recovers the gradient from the least number of machines $f$ theoretically permissible, via an $O(f^2)$ decoding algorithm. We also provide a theoretical delay model which can be used to minimize the expected waiting time per computation by optimally choosing the parameters of the scheme. Finally, we supplement our theoretical findings with numerical results that demonstrate the efficacy of the method and its advantages over competing schemes.
\end{abstract}

\section{Introduction}
With the size of today's datasets, due to high computation and/or memory requirements, it is virtually impossible to run large-scale learning tasks on a single machine; and even if that is possible, the learning process can be extremely slow due to its sequential nature. Therefore, it is highly desirable or, even necessary, to run the tasks in a distributed fashion on multiple machines/cores. For this reason, parallel and distributed computing has attracted a lot of attention in recent years from the machine learning, and other, communities \cite{boyd2011distributed,recht2011hogwild,zinkevich2010parallelized,gemulla2011large}.

When a task is divided among a number of machines, the ``computation time'' is clearly reduced significantly, since the task is being processed in parallel rather than sequentially. However, the taskmaster has to wait for all the machines in order to be able to recover the exact desired computation. Therefore, in the face of substantial or heterogeneous delays, distributed computing may suffer from being slow, which defeats the purpose of the exercise. Several approaches have been proposed to tackle this problem. One naive yet common way, especially when the task consists of many iterations, is to not wait for all machines, and ignore the \emph{straggling machines}. One may hope that in this way on average the taskmaster receives enough information from everyone; however, it is clear that the performance of the learning algorithm may be significantly impacted in many cases because of lost updates. An alternative and more appropriate way to resolve this issue, is to introduce some \emph{redundancy} in the computation of the machines, in order to efficiently trade off computation time for less wait time, and to be able to recover the correct update using only a few machines. But the great challenge here is to design a clever scheme for distributing the task among the machines, such that the computation can be recovered using a few machines, independent of which machines they are.

Over the past few decades, coding theory has been developed to address similar challenges in other domains, and has had enormous success in many applications such as mobile communication, storage, data transmission, and broadcast systems. Despite the existence of a great set of tools developed in coding theory which can be used in many machine learning problems, researchers had not looked at this area until very recently \cite{lee2016speeding,Tandon2016,dutta2016short,li2016fundamental}. This work is aimed at bridging the gap between distributed machine learning and coding theory, by introducing a carefully designed coding scheme for efficiently distributing a learning task among a number of machines. 

More specifically, we consider gradient-based methods for additively separable cost functions, which are the most common way of training any model in machine learning, and use coding to cleverly distribute each gradient iteration across $n$ machines in an efficient way. To that end, we propose a deterministic construction based on Reed-Solomon codes \cite{Reed1960} accompanied with an efficient decoder, which is used to recover the full gradient update from a fixed number of returning machines. Furthermore, we provide a new delay model based on heavy-tail distributions that also incorporates the time required for decoding. We analyze this model theoretically and use it to optimally pick our scheme's parameters. We compare the performance of our method on the MNIST dataset \cite{lecun1998mnist} with other approaches, namely: 1) Ignoring the straggling machines \cite{pan2017revisiting}, 2) Waiting for all the machines, and 3) \textsc{GradientCoding} as proposed by Tandon et al. \cite{Tandon2016}. Our numerical results show that, for the same training time, our scheme achieves better test errors.

\subsection{Related Work}
As mentioned earlier, coding theory in machine learning is a relatively new area. We summarize the recent related work here. Lee et al.~\cite{lee2016speeding} recently employed a coding-theoretic method in two specific distributed tasks, namely matrix multiplication and data shuffling. They showed significant speed-ups are possible in those two tasks by using coding.
Dutta et al.~\cite{dutta2016short} proposed a method that speeds up distributed matrix multiplication by sparsifying the inner products computed at each machine. A coded MapReduce framework was introduced by Li et al in \cite{li2016fundamental} which is used to facilitate data shuffling in distributed computing. The closest work to our framework is the work of Tandon et al.~\cite{Tandon2016}, which aims at mitigating the effect of stragglers in distributed gradient descent using Maximum-Distance Separable (MDS) codes.
However, no analysis of computation time was provided. Furthermore, in their framework, along with the above-mentioned works, the decoding was assumed to be performed offline which might be impractical in certain settings.

\subsection{Statement of Contributions}
In this work, we make the following three main contributions.
\begin{enumerate}
\item We construct a deterministic coding scheme for efficiently distributing gradient descent over a given number of machines. Our scheme is optimal in the sense that it can recover the gradient from the smallest possible number of returning machines, $f$, given a prespecified computational effort per machine.
\item We provide an efficient online decoder, with time complexity $O(f^2)$ for recovering the gradient from any $f$ machines, which is faster than the best known method \cite{Tandon2016}, $O(f^3)$.
\item We analyze the total computation time, and provide a method for finding the optimal coding parameters. We consider heavy-tailed delays, which have been widely observed in CPU job runtimes in practice~\cite{leland1986load,harchol1999ect,harchol1997exploiting}.
\end{enumerate}

The rest of the paper is organized as follow. In Section~\ref{sec:prelim}, we describe the problem setup and explain the design objectives in detail. Section~\ref{sec:construction}, provides the construction of our coding scheme, using the idea of balanced Reed-Solomon codes. Our efficient online decoder is presented in Section~\ref{sec:decoding}. We then characterize the total computation time, and describe the optimal choice of coding parameters, in Section~\ref{sec:time}. Finally, we provide our numerical results in Section~\ref{sec:sim}, and conclude in Section~\ref{sec:conclusion}.
\section{Preliminaries}\label{sec:prelim}
\subsection{Problem Setup}
\begin{figure}[t]
\vskip 0.2in
\begin{center}
\centerline{\includegraphics[width=0.30\columnwidth]{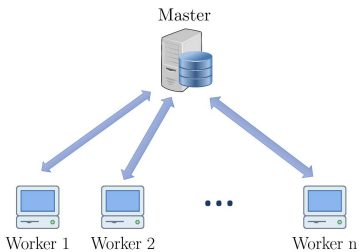}}
\caption{Schematic representation of the taskmaster and the $n$ workers.}
\label{fig:master}
\end{center}
\vskip -0.2in
\end{figure}
Consider a setting where there is a taskmaster $M$ and there are $n$ workers (computing machines) $W_1, W_2,\dots, W_n$ interacting with the taskmaster, as in Fig.~\ref{fig:master}. The master intends to train a model using gradient descent by distributing the gradient updates amongst the workers. More precisely, consider a typical scenario, where we want to learn parameters $\beta \in \RR^p$ by minimizing a generic loss function $L(\mathcal{D};\beta)$ over a given dataset $\mathcal{D} = \{(x_i,y_i)\}_{i = 1}^N$, where $x_i \in \RR^p$ and $y_i \in \RR$. The loss function can be expressed as the sum of the losses for individual data points, i.e. $L(\mathcal{D};\beta) = \sum_{i = 1}^N \ell(x_i,y_i;\beta).$ 
Therefore, the full gradient, with respect to $\beta$, is given by
\begin{equation}
\nabla L(\mathcal{D};\beta) = \sum_{i = 1}^N \nabla \ell(x_i,y_i;\beta). \label{eqn:gradient}
\end{equation}
The data can be divided into $k$ (disjoint) chunks $\{\mathcal{D}_1,\ldots,\mathcal{D}_k\}$ of size $\frac{N}{k}$, and clearly the gradient can also be written as
$\nabla L(\mathcal{D};\beta) = \sum_{i = 1}^{k}\sum_{(x,y) \in \mathcal{D}_i} \nabla \ell(x,y;\beta)$. 
Define $g_i:=\sum_{(x,y) \in \mathcal{D}_i} \nabla \ell(x,y;\beta)$ as the partial gradient of chunk $i$ for every $i$, and $g^\t:=[g_1^\t,g_2^\t,\dots,g_k^\t]$, where $g_i$ is a row vector of length $p$. Therefore $\nabla L(\mathcal{D};\beta)=\mathbf{1}_{1\times k} g$.
Now suppose each worker $W_i$ is assigned $w$ data partitions $\{\mathcal{D}_{i_1},\ldots, \mathcal{D}_{i_w}\}$, on which it computes the partial gradients $\{g_{i_1},g_{i_2},\dots,g_{i_w}\}$. Note that the ``redundancy'' in computation is introduced here, since each chunk is allowed to be assigned to multiple workers. Each worker then has to compute its partial gradients, and return a prespecified \emph{linear combination} of them to the master.

As it will be explained in detail, $k$ and $w$ are to be chosen in such a way that the total computation time is minimized. For a fixed $k$ and $w$, we want to be able to recover the gradient using the linear combinations received from the fastest $f$ machines at the master (or equivalently tolerate $s:=n-f$ stragglers). 
Note that we do not assume any prior knowledge about the stragglers, i.e., we shall design a scheme that enables master to recover the gradient from any set of $f$ machines.
%
%
%
%
%
It is known \cite{Tandon2016} that for any fixed $k$ and $w$, an upper-bound on the number of stragglers that \emph{any scheme} can tolerate is:
\begin{equation}
s\leq \floor{\frac{w n}{k}}-1. \label{eqn:stragglers_bound}
\end{equation}
The scheme proposed in this work achieves this bound. 
A coding scheme designed to tolerate $s$ stragglers consists of an $\emph{encoding matrix}$ $\B$, and a collection of decoding vectors $\{\mathbf{a}_\mathcal{F} : \mathcal{F} \subset [n], \card{\mathcal{F}} = n - s\}$. The matrix $\B$ should satisfy:
\begin{enumerate}
\item Each row of $\B$ contains exactly $w$ nonzero entries.
\item The linear space generated by any $f$ rows of $\B$ contains the all-one vector of length $k$,  $\mathbf{1}_{1 \times k}$.
\end{enumerate}
The values of these nonzero entries prescribe the linear combination sent by $W_i$. In other words, the \emph{coded partial gradient} sent from $W_i$ to $M$ is given by
\begin{equation}
c_i = \sum_{j = 1}^{k}\B_{i,j} g_j = \B_ig,
\end{equation}
where $\mathbf{B}_i$ denotes the $i^\text{th}$ row of $\B$. The $c_i\text{'s}$ define the encoded computation matrix $\mathbf{C} \in \CC^{n\times p}$ as $\mathbf{C}  = \begin{bmatrix}c_1^\t & c_2^\t & \cdots & c_n^\t \end{bmatrix}^\t= \B g$, where $c_i \in \CC^{1 \times p}$.
The decoding vectors are chosen as follows: let $\mathcal{F} = \{i_1,\ldots,i_f\}$ be the indices of the returning machines and let $\B_\mathcal{F}$ be the sub-matrix of $\B$ with rows indexed by $\mathcal{F}$. If $\mathbf 1_{1\times k}$ is in the linear space generated by the rows of $\mathbf{B}_{\mathcal F}$, as the second property suggests, $\mathbf{a}_{\mathcal F}$ is chosen such that $\mathbf{a}_{\mathcal F} \mathbf{B}_{\mathcal F}=\mathbf 1_{1\times k}$. As a result, we have
\begin{equation}
\label{eqn:gradient_survivors}
\mathbf{a}_{\mathcal F}\mathbf{C}_{\mathcal F}=\mathbf{a}_{\mathcal F}\mathbf{B}_{\mathcal F}g=\mathbf 1_{1\times k} \; g=\nabla L(\mathcal{D};\beta) .
\end{equation}
When this holds for any set of indices $\mathcal{F} \subset [n]$ of size $f$, it means that the gradient can be recovered from the set of $f$ machines that return fastest. 
\cite{Tandon2016},
\subsection{Computational Trade-offs}
In a distributed scheme that does not employ redundancy, the taskmaster has to wait for all the workers to finish in order to compute the full gradient. However, in the scheme outlined above, the taskmaster needs to wait for the fastest $f$ machines to recover the full gradient. Clearly, this requires more computation by each machine.
Note that in the uncoded setting, the amount of computation that each worker does is $\frac{1}{n}$ of the total work, whereas in the coded setting each machine performs a $\frac{w}{k}$ fraction of the total work. From~\eqref{eqn:stragglers_bound}, we know that if a scheme can tolerate $s$ stragglers, the fraction of computation that each worker does is
$\frac{w}{k}\geq \frac{s+1}{n}$.
Therefore, the computation load of each worker increases by a factor of $(s+1)$.
As will be explained further in Section~\ref{sec:time}, there is a sweet spot for $\frac{w}{k}$ (and consequently $s$) that minimizes the expected total time that the master waits in order to recover the full gradient update.

It is worth noting that it is often assumed \cite{Tandon2016,lee2016speeding,dutta2016short} that the decoding vectors are precomputed for all possible combinations of returning machines, and the decoding cost is not taken into account in the total computation time. In a practical system, however, it is not very reasonable to compute and store all the decoding vectors, especially as there are ${n \choose f}$ such vectors, which grows quickly with $n$. In this work, we introduce an online algorithm for computing the decoding vectors on the fly, for the indices of the $f$ workers that respond first. The approach is based on the idea of inverting Vandermonde matrices, which can be done very efficiently. 
In the sequel, we show how to construct an encoding matrix $\B$ for any $w,k$ and $n$, such that the system is resilient to $\floor{\frac{w n}{k}} - 1$ stragglers, along with an efficient algorithm for computing the decoding vectors $\{\mathbf{a}_\mathcal{F}: \mathcal{F} \subset [n], \card{\mathcal{F}} = f\}$.
\vspace{-0.2in}
\section{Code Construction}\label{sec:construction}
\renewcommand{\algorithmicrequire}{\textbf{Input:}}
\renewcommand{\algorithmicensure}{\textbf{Output:}}
The basic building block of our encoding scheme is a matrix $\M \in \{0,1\}^{n \times k}$, where each row is of weight $w$, which serves as a $\emph{mask}$ for the matrix $\B$, where $w$ is the number of data partitions that is assigned to every machine. Each column of $\B$ will be chosen as a codeword from a suitable Reed--Solomon Code over the complex field, with support dictated by the corresponding column in $\M$. Whereas the authors of~\cite{Tandon2016} choose the \emph{rows} of $\B$ as codewords from a suitable MDS code, this approach does not immediately work when $k$ is not equal to $n$.
\subsection{Balanced Mask Matrices}
We will utilize techniques from~\cite{Halbawi2016ISIT, Halbawi2016ITW} to construct the matrix $\M$ (and then $\B$). For that, we present the following definition.
\begin{defn}[Balanced Matrix]
A matrix $\M \in \{0,1\}^{n \times k}$ is column (row)-balanced if for fixed row (column) weight, the weights of any two columns (rows) differ by at most 1.
\end{defn}
\begin{rem}
While our scheme can handle the case where $\frac{nw}{k}$ is not an integer, in this section we illustrate the case where it is. The general result is described in the appendix.
\end{rem}
Ultimately, we are interested in a matrix $\M$ with row weight $w$ that prescribes a mask for the encoding matrix $\B$. As an example, let $n = 8$, $k = 4$ and $w = 3$. Then, $\M$ is given by
\begin{equation}
\M = \begin{bmatrix}
1 & 1 & 1 & 0\\
1 & 1 & 1 & 0\\
1 & 1 & 0 & 1\\
1 & 1 & 0 & 1\\
1 & 0 & 1 & 1\\
1 & 0 & 1 & 1\\
0 & 1 & 1 & 1\\
0 & 1 & 1 & 1
\end{bmatrix}, \label{eqn:mask_example}
\end{equation}
where each column is of weight $\frac{nw}{k} = 6$. 
The following algorithm produces a balanced mask matrix. For a fixed column weight $d$, each row has weight either $\floor{\frac{kd}{n}}$ or $\ceil{\frac{kd}{n}}$.
\begin{algorithm}[H]
\caption{\textsc{RowBalancedMaskMatrix}($n$,$k$,$d$,$t$)}
\begin{algorithmic}
\Require \\
	     $n$: Number of rows, 
		 $k$: Number of columns, 
		 $d$: Weight of each column, 
		 $t$: Offset parameter
\Ensure Row-balanced $\M \in \{0,1\}^{n\times k}$
\State $\M \gets \mathbf{0}_{n\times k}$
\For{$j = 0$ to $k-1$}
	\For{$i = 0$ to $d - 1$}
		\State $r = (i+ jd + t)_n$ \hfill The quantity $(x)_n$ denotes $x$ modulo $n$.
		\State $M_{r,j} = 1$
	\EndFor 
\EndFor
\State \Return{$\M$}
\end{algorithmic}
\label{alg:RB_mask_matrix}
\end{algorithm}
\vspace{-0.2in}
As a result, when $d$ is chosen as $\frac{nw}{k} \in \ZZ$, all rows will be of weight $w$. 
As an example, the matrix $\M$ in~\eqref{eqn:mask_example} is generated by calling \textsc{RowBalancedMaskMatrix}(8,4,6,0). Algorithm~\ref{alg:RB_mask_matrix} can be used to generate a mask matrix $\M$ for the encoding matrix $\B$: The $j^\text{th}$ column of $\B$ will be chosen as a Reed--Solomon codeword whose support is that of the $j^\text{th}$ column of $\M$. 
\vspace{-0.1in}
\subsection{Reed--Solomon Codes}
This subsection provides a quick overview of Reed--Solomon Codes. A Reed--Solomon code of length $n$ and dimension $f$ is a linear subspace $\RS{n}{f}$ of $\CC^n$ corresponding to the evaluation of polynomials of degree less than $f$ with coefficients in $\CC$ on a set of $n$ distinct points $\{\alpha_1,\ldots,\alpha_n\}$, also chosen from $\CC$. When $\alpha_i = \alpha^i$, where $\alpha \in \CC$ is an $n^\text{th}$ root of unity, the evaluations of the polynomial $t(x) = \sum_{i = 0}^{f - 1}t_ix^i$ on $\{1,\alpha,\ldots, \alpha^{n -1}\}$ corresponds to 
\begin{equation}
\begin{bmatrix}
t(1)\\
t(\alpha)\\
\vdots\\
t(\alpha^{n-1})
\end{bmatrix}
= 
\mathbf{G} \mathbf{t}=
\begin{bmatrix}
1 & 1 & \cdots & 1\\
1 & \alpha & \cdots & \alpha^{f - 1}\\
\vdots & \vdots & \ddots & \vdots\\
1 & \alpha^{n - 1} & \cdots & \alpha^{(n - 1)(f - 1)}\\
\end{bmatrix}
\begin{bmatrix}
t_0\\
t_1\\
\vdots\\
t_{f - 1}
\end{bmatrix} \label{eqn:polynomial_evaluation}.
\end{equation}
It is well-known that any $f$ rows of $\G$ form an invertible matrix, which implies that specifying any $f$ evaluations $\{t(\alpha^{i_1}),\ldots, t(\alpha^{i_f})\}$ of a polynomial $t(x)$ of degree at most $f-1$ characterizes it. In particular, fixing $f-1$ evaluations of the polynomial to zero characterizes $t(x)$ uniquely up to scaling. This property will give us the ability to construct $\B$ from $\M$.
\vspace{-0.1in}
\subsection{Building the Encoding Matrix from the Mask Matrix}
Once a mask matrix $\M$ has been determined using Algorithm \ref{alg:CB_mask_matrix}, the encoding matrix $\B$ can be built by picking appropriate codewords from $\RS{n}{f}$. Consider $\M$ in \eqref{eqn:mask_example} and the following polynomials
\begin{eqnarray}
t_1(x) &=& \kappa_1(x - \alpha^6)(x - \alpha^7),\\
t_2(x) &=& \kappa_2(x - \alpha^4)(x - \alpha^5),\\
t_3(x) &=& \kappa_3(x - \alpha^2)(x - \alpha^3),\\
t_4(x) &=& \kappa_4(x - 1)(x - \alpha).
\end{eqnarray}
The constant $\kappa_j$ is chosen such that the constant term of $t_j(x)$, i.e. $t_j(0)$, is equal to $1$. The evaluations of $t_j(x)$ on $\{1,\alpha,\ldots,\alpha^7\}$ are collected in the vector $(t_j(1),t_j(\alpha),\ldots,t_j(\alpha^7))^\t$ which sits as the $j^\text{th}$ column of $\B$. The validity of this process can be confirmed using~\eqref{eqn:polynomial_evaluation}, and is generalized in Algorithm~\ref{alg:encoding}.
\vspace{-0.1in}
\begin{algorithm}[H]
\caption{\textsc{EncodingMatrix}($n$,$k$,$w$)}
\label{alg:encoding}
\begin{algorithmic}
\Require \\
	     $n$: Number of rows, 
		 $k$: Number of columns, 
		 $w$: Row weight, 
		 $\alpha$: $n^\text{th}$ root of unity
\Ensure Row-balanced encoding matrix $\B$.
\State $\M \gets \Call{MaskMatrix}{n,k,w}$
\State $\B \gets \mathbf{0}_{n \times k}$
\For{$j = 0$ to $k-1$}
	\State $t_j(x) \gets \prod_{r: \M_{r,j} = 0}(x - \alpha^r)/(-\alpha^r)$
	\For{$i = 0$ to $n - 1$}
		\State $\B_{i,j} = t_j(\alpha^i)$
	\EndFor 
\EndFor
\State \Return{$\B$}
\end{algorithmic}
\end{algorithm}
\vspace{-0.2in}
Once the matrix $\B$ is specified, the corresponding decoding vectors required for computing the gradient at the taskmaster have to be characterized.
\section{Efficient Online Decoding}\label{sec:decoding}
We exploit the fact that $\B$ is constructed using Reed--Solomon codewords and show that each decoding vector $\mathbf{a}_\mathcal{F}$ can be computed in $O(f^2)$ time. Recall that the taskmaster should be able to compute the gradient from \emph{any} $f$ surviving machines, indexed by $\mathcal{F} \subseteq [n]$, according to \eqref{eqn:gradient_survivors}.
The $j^\text{th}$ column of $\B$ is determined by a polynomial $t_j(x) = \sum_{i = 0}^{f - 1}t_{j,i}x^i$ where $t_{j,0} = 1$.
We can write $\B$ as $\B = \G \T$, where $\T = \begin{bmatrix} \mathbf{t}_1 & \cdots & \mathbf{t}_k\end{bmatrix}$ and $\mathbf{t}_j$ is the vector of coefficients of $t_j(x)$, and $\G$ is the matrix given in~\eqref{eqn:polynomial_evaluation}.
Now consider $\mathbf{C}_\mathcal{F}$, the coded partial gradients received from $\{W_i : i \in \mathcal{F}\}$. The rows of $\B$ corresponding to $\mathcal{F}$ are given by
\begin{equation}
\B_\mathcal{F} = \G_\mathcal{F} \T = \begin{bmatrix}
1 & \alpha^{i_1} & \cdots & \alpha^{i_1(f - 1)}\\
1 & \alpha^{i_2} & \cdots & \alpha^{i_2(f - 1)}\\
\vdots & \vdots & \ddots & \vdots\\
1 & \alpha^{i_f} & \cdots & \alpha^{i_f(f - 1)}\\
\end{bmatrix}
\begin{bmatrix}
1 & \cdots & 1\\
t_{1,1} & \cdots & t_{k,1}\\
\vdots  & \ddots & \vdots\\
t_{1,f-1} & \cdots & t_{k,f-1}.
\end{bmatrix}.
\end{equation}
We require a vector $\mathbf{a}_\mathcal{F}$ such that $\mathbf{a}^\t_\mathcal{F} \B_\mathcal{F} = \mathbf{1}_{1 \times k}$. This is equivalent to finding a vector $\mathbf{a}_\mathcal{F}$ such that 
\begin{equation}
\mathbf{a}_\mathcal{F}^\t \G_\mathcal{F}= (1,0,\ldots,0).
\end{equation}
Indeed, the matrix $\G_\mathcal{F}$ in the above product is a Vandermonde matrix defined by $f$ distinct elements and so it is invertible in $O(f^2)$ time~\cite{bjorck1970solution}, which facilitates the online computation of the decoding vectors. 
This is an improvement compared to previous works~\cite{Tandon2016} where the decoding time is usually $O(f^3)$.
A careful inspection of inverses of Vandermonde matrices built from an $n^\text{th}$ root of unity allows us to compute the required decoding vector in a space efficient manner. This is demonstrated in the next subsection.
\subsection{Space-Efficient Algorithm}
Note that $\mathbf{a}_\mathcal{F}^\t$ is nothing but the first row of the inverse of $\G_\mathcal{F}$, which can be built from a set of polynomials $\{v_1(x),\ldots,v_f(x)\}$. Let the $l^\text{th}$ column of $\G^{-1}_\mathcal{F}$ be $\mathbf{v}_l = (v_{l,0},\ldots,v_{l,f - 1})^\t$ and associate it with $v_l(x) =\sum_{i = 0}^{f - 1}v_{l,i}x^i$. The condition $\G_\mathcal{F} \mathbf{v}_l = \mathbf{e}_l$, where $\mathbf{e}_l$ is the $l^\text{th}$ elementary basis vector of length $f$, implies that $v_l(x)$ should vanish on $\{\alpha^{i_1},\ldots,\alpha^{i_f}\}\setminus\{\alpha^{i_l}\}$. Specifically,
\begin{equation}
v_l(x) = \prod_{\substack{j = 1 \\ j \neq l}}^{f} \frac{x - \alpha^{i_j}}{\alpha^{i_l} - \alpha^{i_j}}
\end{equation}
The first row of $\G^{-1}_\mathcal{F}$ is given by $(v_{1,0},\ldots,v_{f,0})$, where $v_{l,0}$ is the constant term of $v_l(x)$. Indeed, we have $v_{l,0} = v_l(0)$, which can be computed in closed form according to the following formula,
\begin{equation}
v_{l,0} = \prod_{\substack{j = 1 \\ j \neq l}}^{f} \frac{\alpha^{i_j}}{ \alpha^{i_j}-\alpha^{i_l} } = \prod_{\substack{j = 1 \\ j \neq l}}^{f} (1 - \alpha^{i_l - i_j})^{-1}.
\end{equation}
By choosing $\alpha$ as a primitive $n^\text{th}$ root of unity, one is guaranteed that there are only $n - 1$ distinct values of $(1 - \alpha^{i_l - i_j})^{-1}$. This observation proposes that the master should precompute and store the set $\{(1 - \alpha^i)^{-1}\}_{i= 1}^{n - 1}$, and then compute each $v_{l,0}$ by utilizing lookup operations. The following algorithm outlines this procedure.
\vspace{-0.1in}
\begin{algorithm}[H]
\caption{\textsc{DecodingVector}($\mathcal{F}$)}
\label{alg:decoding}
\begin{algorithmic}
\Require \\
	     $\mathcal{F}$: Ordered set of surviving machines - $\{i_1, \ldots i_{f}\}$\\
		 $\alpha$: $n^\text{th}$ root of unity
\Ensure Decoding vector $\mathbf{a}$ associated with $\mathcal{F}$.
\State $\mathbf{a} \gets \mathbf{0}_{f}$
\For{$ l= 1$ to $f$}
\State $\mathbf{a}_{l} \gets	\prod_{j = 0, j \neq l}^{f-1} (1 - \alpha^{i_l - i_j})^{-1}$
\EndFor
\State \Return{$\mathbf{a}$}
\end{algorithmic}
\end{algorithm}
\vspace{-0.2in}
\section{Analysis of Total Computation time}\label{sec:time}
In this section, we provide a theoretical model which can be used to optimize the choice of parameters that define the encoding scheme. For this purpose, we model the response time of a single computing machine as 
\begin{equation}
T = T_{\text{delay}}+T_{\text{comp}}.
\end{equation}
Here, the quantity $T_{\text{comp}}$ is the time required for a machine to compute its portion of the gradient. 
This quantity is equal to $c_g\frac{Nw}{k}$, where $c_g=c_g(\ell, p)$ is a constant that indicates the time of computing the gradient for a single data point which depends on the dimension of data points, $p$, as well as the loss function, $\ell$.
The second term $T_\text{delay}$ reflects the random delay incurred before the machine returns with the result of its computation. We model this delay as a Pareto distributed random variable with distribution function
\begin{equation}
F(t) = Pr(T_\text{delay}\leq t)=1-\left(\frac{t_0}{t}\right)^{\xi}\;\;\;\text{for}\;\; t\geq t_0, \label{delay:eqn:pareto}
\end{equation}
where the quantity $t_0$ can be thought of the fundamental delay of the machine, i.e. the minimum time required for a machine to return in perfect conditions. Previous works~\cite{lee2016speeding, liang2014tofec} model the return time of a machine as a shifted exponential random variable. We propose using this approach since the heavy-tailed nature of CPU job runtime has been observed in practice~\cite{leland1986load,harchol1999ect,harchol1997exploiting}.

Let $T_f$ denote the expected time of computing the gradient using the first $f$ machines. As a result we have 
\begin{equation}
T_f = \mathbb E[T_\text{delay}^{(f)}]+T_{\text{comp}}+T_{\text{dec}}(f), \label{delay:eqn:model_dec}
\end{equation}
where $T_\text{delay}^{(f)}$ is the $f^\text{th}$ ordered statistic of $T_\text{delay}$, and $T_\text{dec}(f)$ is the time required at the taskmaster for decoding. Here we assume $n$ is large and define $\alpha:=\frac{w}{k}$ as the fraction of the dataset assigned to each machine. For this value of $\alpha$, the number of machines required for successful recovery of the gradient is given by
\begin{equation}
f(\alpha) = \ceil{(1 - \alpha) n} + 1, \label{delay:eqn:f(a)}
\end{equation}
where $\ceil{x}$ returns the smallest integer greater than or equal to $x$. We can show the following result which approximates $\mathbb E[T_{\text{delay}}^{(f)}]$ for large values of $n$. For a proof, please refer to the appendix.
\begin{prop}
\label{prop2}
The expected value of the $f^\text{th}$ order statistic of the Pareto distribution with parameter $\xi$ will converge as $n$ grows, i.e.,
\begin{equation}
 \lim_{n \rightarrow \infty} \mathbb E[T_\text{delay}^{(f)}]= \lim_{n \rightarrow \infty} \mathbb E[T_\text{delay}^{(1-\alpha) n}]= t_0\alpha^{-\frac{1}{\xi}}.
\end{equation}
\end{prop}
Using this result, we can approximate $T_f$, for $n\gg 1$,
\begin{equation}
\label{eqn:total_time}
T_f\approx t_0 \alpha^{-\frac{1}{\xi}}+ c_g N \alpha+c_m(1-\alpha)^2 n^2,
\end{equation}
where we assume that the taskmaster uses Algorithm \ref{alg:decoding} for decoding. 
If we assume $c_m$ is the time required for one FLOP, the total decoding time is given by $c_m(f-1)f\approx c_m(1-\alpha)^2 n^2$. Since $\alpha$ is bounded from above by the memory of each machine, one can find the optimal computation time, subject to memory constraints, by minimizing $T_f$ with respect to $\alpha$.

\subsection{Offline Decoding}
In the schemes where the decoding vectors are computed offline, the quantity $T_\text{dec}$ does not appear in the total computation time $T_f$. Therefore, for large values of $n$, we can write:

\begin{equation}
T_f=t_0 \alpha^{-\frac{1}{\xi}}+ c_g N \alpha. \label{delay:model_asymp}
\end{equation}
This function can be minimized with respect to $\alpha$ by standard calculus to give 
\begin{equation}
\label{eqn:optimal_alpha}
\alpha^* = \left(\frac{t_0}{c_g N\xi}\right)^\frac{\xi}{1 + \xi}.
\end{equation}
Note that this quantity is valid (less than one) if and only if one has $\frac{t_0}{c_gN \xi} < 1$. It has been observed in practice that the parameter $\xi$ is close to one. Therefore, this assumption holds because $N$ is assumed to be large. 
\section{Numerical Results}\label{sec:sim}
\begin{SCfigure}[][h]
\includegraphics[width=0.5\textwidth]{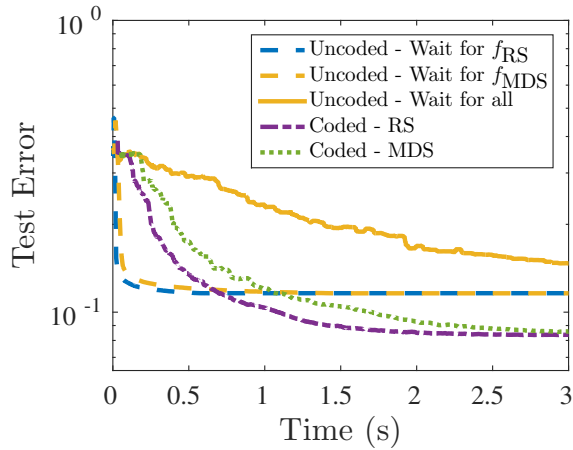}
\caption{This plot shows the test error as a function of time for a softmax regression model trained using distributed gradient descent. The model was trained on $n = 80$ machines using 12000 examples from the MNIST database~\cite{lecun1998mnist} and was validated on a test set of size 10000. The Reed--Solomon based scheme (Coded - RS) waits for $f_{\text{RS}} = 68$ machines, while the one corresponding to~\cite{Tandon2016} (Coded - MDS) waits for $f_{\text{MDS}} = 33$. These quantities were obtained by numerically optimizing~\eqref{delay:eqn:model_dec}. For comparison, we plot the performance of three uncoded schemes: One waits for all 80 machines, another waits for $f_\text{RS}$ machines and the third waits for $f_\text{MDS}$ of them.}
\label{fig:exp}
\end{SCfigure}

To demonstrate the effectiveness of the scheme, we performed numerical simulations using MATLAB. We train a softmax regression model on a distributed cluster composed of $n = 80$ machines to classify 10000 handwritten digits from the MNIST database~\cite{lecun1998mnist} while artificially introducing delay as a random variable sampled from a Pareto distribution according to~\eqref{delay:eqn:pareto} with parameters $\xi = 1.1$ and $t_0 = 0.001$. 
Similar to~\cite{Tandon2016}, knowledge of the entire gradient allows us to employ accelerated gradient methods such as the one proposed by Nesterov~\cite{nesterov2013introductory}. Details of the  experiment are given in the accompanying description of Figure~\ref{fig:exp}.
We compare several schemes by running each of them on the same dataset for a fixed amount of time (in seconds) and then measuring the test error. The results depicted in Figure~\ref{fig:exp} demonstrate that the scheme proposed in this paper outperforms the four other schemes.
\vspace{-0.3in}
\section{Conclusion}\label{sec:conclusion}
We presented  a straggler mitigation scheme that facilitates the implementation of distributed gradient descent in a computing cluster. For a fixed per-machine computational effort, the taskmaster recovers the full gradient from the least number of machines theoretically required, which is done via an algorithm that is efficient in both space and time. Furthermore, we propose a theoretical delay model based on heavy-tailed distributions and incorporates the decoding time, which allows us to minimize the expected running time of the algorithm.
\pagebreak
\bibliography{library}
\bibliographystyle{plain}
\newpage
\section{Appendix}
\subsection{Correctness of Algorithm \ref{alg:RB_mask_matrix}}
To lighten notation, we prove correctness for $t = 0$. The general case follows immediately.
\begin{prop}
\label{prop:balanced_matrix}
Let $k,d$ and $n$ be integers where $d < n$. The row weights of matrix $\M \in \{0,1\}^{n \times k}$ produced by Algorithm~\ref{alg:RB_mask_matrix} for $t = 0$ are
\begin{equation}
w_i = \begin{dcases}
\ceil{\frac{kd}{n}},& i \in \{0, \ldots, (kd - 1)_n\}, \\
\floor{\frac{kd}{n}},& i \in \{(kd)_n,\ldots, n - 1\}.
\end{dcases}
\end{equation}
\end{prop}
\textit{Proof.} The nonzero entries in column $j$ of $\M$ are given by 
\begin{equation}
\mathcal{S}_j = \{jd,\ldots, (j+1)d - 1\}_n,
\end{equation}
where the subscript $n$ denotes reducing the elements of the set modulo $n$. Collectively, the nonzero indices in all columns are given by
\begin{equation}
\mathcal{S} = \{0, \ldots d - 1,\ldots, (k - 1)d,\ldots kd - 1\}_n.
\end{equation}

In case $n \mid kd$, each element in $\mathcal{S}$, after reducing modulo $n$, appears the same number of times. As a result, those indices correspond to columns of equal weight, namely $\frac{kd}{n}$. Hence, the two cases of $w_i$ are identical along with their corresponding index sets.

In the case where $n \nmid kd$, each of the first $\floor{\frac{kd}{n}}n$ elements, after reducing modulo $n$, appears the same number of times. As a result, the nonzero entries corresponding to those indices are distributed evenly amongst the $n$ rows, each of which is of weight $\floor{\frac{kd}{n}}$. The remaining indices $\{\floor{\frac{kd}{n}}n,\ldots, kd - 1\}_n$ contribute an additional nonzero entry to their respective rows, those indexed by $\{0,\ldots, (kd - 1)_n\}$. Finally, we have that the first $(kd)_n$ rows are of weight $\floor{\frac{kd}{n}} + 1 = \ceil{\frac{kd}{n}}$, while the remaining ones are of weight $\floor{\frac{kd}{n}}$. $\square$

Now consider the case when $t$ is not necessarily equal to zero. This amounts to shifting (cyclically) the entries in each column by $t$ positions downwards. As a result, the rows themselves are shifted by the same amount allowing to conclude us the following.
\begin{clry}
\label{clry:balanced_matrix_offset}
Let $k,d$ and $n$ be integers where $d < n$. The row weights of matrix $M \in \{0,1\}^{n \times k}$ produced by Algorithm~\ref{alg:RB_mask_matrix} are
\begin{equation}
w_i = \begin{dcases}
\ceil{\frac{kd}{n}}& i \in \{t, \ldots, (t + kd - 1)_n\}, \\
\floor{\frac{kd}{n}}& i \in \{0, t - 1\} \cup \{(t + kd)_n,\ldots, n - 1\}.
\end{dcases}
\end{equation}
\end{clry}

\subsection{General construction}
In case $d = \frac{w k}{n} \notin \mathbb{Z}$, the chosen row weight $w$ prevents the existence of $\M$ where each column weight is minimal. We have to resort to Algorithm \ref{alg:CB_mask_matrix} that yields $\M$ comprised of two matrices $\M_h$ and $\M_l$ according to
\begin{equation}
\M =
\begin{bmatrix}
\M_h & \M_l
\end{bmatrix}.
\end{equation}
The matrices $\M_h$ and $\M_l$ are constructed using Algorithm \ref{alg:RB_mask_matrix}. Each column of $\M_h$ has weight $d_h := \ceil{\frac{nw}{k}}$ and each column of $\M_l$ has weight $d_l := \floor{\frac{nw}{k}}$. Note that according to \eqref{eqn:stragglers_bound}, we require $d_l \geq 2$ in order to tolerate a positive number of stragglers.
\begin{algorithm}[H]
\caption{Column-balanced Mask Matrix $\M$}
\begin{algorithmic}
\Require \\
	     $n$: Number of rows\\
		 $k$: Number of columns\\
		 $w$: Weight of each row
\Ensure Row-balanced $\M \in \{0,1\}^{n,k}$.
\Procedure{MaskMatrix}{$n$,$k$,$w$}
\State $k_h \gets (nw)_k$
\State $d_h \gets \ceil{\frac{nw}{k}}$
\State $k_l \gets k - k_h$
\State $d_l \gets \floor{\frac{nw}{k}}$
\State $t \gets (k_h d_h)_n$
\State $\M_h \gets \Call{RowBalancedMaskMatrix}{n,k_h,d_h, 0}$
\State $\M_l \gets \Call{RowBalancedMaskMatrix}{n,k_l,d_l, t}$
\State $\M \gets \begin{bmatrix} \M_h & \M_l\end{bmatrix}$
\State \Return{$\M$}
\EndProcedure
\end{algorithmic}
\label{alg:CB_mask_matrix}
\end{algorithm}
\subsection{Correctness of Algorithm~\ref{alg:CB_mask_matrix}}
According to the algorithm, the condition $k \mid nw$ implies that $k_h = 0$ leading to $M = M_l$, which is constructed using Algorithm~\ref{alg:RB_mask_matrix}.

Moving on to the general case, the matrix $\M$ given by
\begin{equation}
\begin{bmatrix}
\M_h & \M_l
\end{bmatrix}
\end{equation}
where each matrix is row-balanced. The particular choice of $t$ in $\M_l$ aligns the ``heavy'' rows of $\M_h$ with the ``light" rows of $\M_l$, and vice-versa. The algorithm works because the choice of parameters equates the number of heavy rows $n_h$ of $\M_l$ to the number of light rows $n_l$ of $\M_h$. The following lemma is useful in two ways.
\begin{lem}
\label{lem:weights}
$\floor{\frac{k_h d_h}{n}} + \ceil{\frac{k_l d_l}{n}} = \ceil{\frac{k_h d_h}{n}} + \floor{\frac{k_l d_l}{n}} = w$.
\end{lem}
\textit{Proof.} Note that the following holds:
\begin{equation}
\frac{k_h d_h}{n} + \frac{k_l d_l}{n} - 1 <  \ceil{\frac{k_h d_h}{n}} + \floor{\frac{k_l d_l}{n}}  < \frac{k_h d_h}{n} + \frac{k_l d_l}{n} + 1.
\end{equation}
Furthermore, we have that 
\begin{eqnarray}
\frac{k_h d_h}{n} + \frac{k_l d_l}{n} &=& \frac{k_h(d_l + 1)}{n} + \frac{(k - k_h)d_l}{n}\\
									  &=& \frac{k_h}{n} + \frac{k d_l}{n}\\
									  &=& \frac{wn - \floor{\frac{wn}{k}}k}{n} + \frac{k d_l}{n}\\
									  &=&\frac{wn - d_lk}{n} +  \frac{k d_l}{n}\\
									  &=& w.
\end{eqnarray}
We combine the two observations in one:
\begin{equation}
w - 1 <  \ceil{\frac{k_h d_h}{n}} + \floor{\frac{k_l d_l}{n}}  < w + 1,
\end{equation}
and conclude that $\ceil{\frac{k_h d_h}{n}} + \floor{\frac{k_l d_l}{n}} = w$. $\square$

We have shown the concatenation of a ``heavy'' row of $\M_h$ along with a ``light'' row of $\M_l$ results in one that is of weight $w$. It remains to show that the concatenation of $\M_h$ and $\M_l$ results of rows of this type only.

We will assume that $n \nmid k_h d_h$ holds. From Proposition~\ref{prop:balanced_matrix}, we have $n_l = n - (k_h d_h)_n$ and $n_h = (k_l d_l)_n$. We will show that the two quantities are in fact equal.
Indeed, we can express $n_l$ as 
\begin{eqnarray}
n- (k_h d_h)_n &=& n - k_h d_h + \floor{\frac{k_h d_h}{n}} n\\
    &=& - k_h d_h + \ceil{\frac{k_h d_h}{n}} n\\
	&=& -(k - k_l)(d_l + 1) + \ceil{\frac{k_h d_h}{n}}n\\
	&=& -(d_l k + k_h) + k_l d_l - \floor{\frac{k_l d_l}{n}} n + nw\\
	&=& k_l d_l - \floor{\frac{k_l d_l}{n}}n\\
	&=& (k_l d_l)_n
\end{eqnarray}
Hence $n_l = n_h$ and by the choice of $t$, the ``light" rows of $\M_h$ align with the ``heavy" rows of $\M_l$, and vice-versa. Furthermore, Lemma~\ref{lem:weights} guarantees that each row of $\M$ is of weight $\ceil{\frac{k_h d_h}{n}} + \floor{\frac{k_l d_l}{n}} = w$. The same holds for the remaining rows, using the fact that $\ceil{x} + \floor{y} = \floor{x} + \ceil{y}$ when both $x$ and $y$ are non-integers. 

\subsection{Proof of Proposition \ref{prop2}}

From~\cite{vannman1976estimators}, the expected value of the $f^\text{th}$ ordered statistic of the Pareto distribution is:
\begin{equation*}
\mathbb E[T_\text{delay}^{(f)}] = t_0\frac{\Gamma(n - f + 1 - 1/\xi)\Gamma(n + 1)}{\Gamma(n - f + 1)\Gamma(n + 1 - 1/\xi)},
\end{equation*}
where $\Gamma(x)$ is the gamma function given by $\Gamma(x) = \int_{0}^{\infty}t^{x - 1}e^{-t}dt$. We now assume that $n$ is large and make the standard approximation 
\begin{equation*}
\Gamma(x) \sim \sqrt{\frac{2 \pi}{x}}\left(\frac{x}{e}\right)^x.
\end{equation*}
Furthermore, \eqref{eqn:stragglers_bound} implies that the number of machines we wait for is $f = (1-\alpha) n$, for some $\alpha < 1$ which leads to
\begin{equation*}
\mathbb E[T_\text{delay}^{(f)}] = t_0 \left(1 - \frac{1}{\xi (\alpha n + 1)}\right)^{\alpha n+ \frac{1}{2}} \left(1 - \frac{1}{\xi (n + 1)}\right)^{-n- \frac{1}{2}}\left(1 - \frac{(1-\alpha) n}{n + 1 - \frac{1}{\xi}}\right)^{-1/\xi}.
\end{equation*}
By letting $n \rightarrow \infty$, the first two terms in the product converge to $e^-\xi$ and $e^{\xi}$, respectively, which yields
\begin{equation*}
\lim_{n \rightarrow \infty} \mathbb E[T_\text{delay}^{(f)}]= t_0\alpha^{-\frac{1}{\xi}}.
\end{equation*}

\section{Offline Decoding}
\begin{figure}[H]
\centering
\includegraphics[scale=0.5]{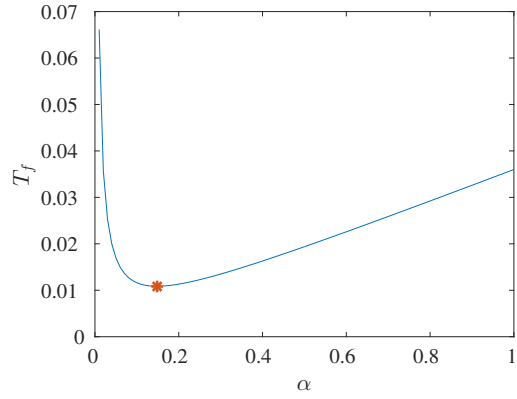}
\caption{This plot corresponds to a setup where the number of training examples is $N = 12000$ and $c_g = 3\times 10^{-6}$ to give $Nc_g = 0.035$. The parameters of the Pareto distribution corresponding to the delay is characterized by $t_0 = 0.001$ and $\xi = 1.1$. The optimizer of this function as predicted by \eqref{eqn:optimal_alpha} is $\alpha^* = 0.1477$. This point is denote by the star symbol.}
\label{fig:delay_theory}
\end{figure}

For illustrative purposes, we plot the function $T_f$ from~\eqref{delay:model_asymp} for a given set of parameters and indicate the optimal point. This plot is given in Figure~\ref{fig:delay_theory}.

\end{document}